\documentstyle[preprint,prl,aps]{revtex}

\title{
The analysis of particle multiplicities in Pb+Pb collisions \\
at 158$A$ GeV/$c$ within hadron gas models
}

\author{
Granddon D. Yen$^1$ and
Mark I. Gorenstein$^{2,3,}$\footnote{Permanent address: 
Bogolyubov Institute for Theoretical Physics, Kiev, Ukraine}
}

\address{
$^1$Institute of Physics, Academia Sinica, Taipei 11529, Taiwan \\
 $^2$ School of Physics and Astronomy, Tel Aviv University, 
Tel Aviv, Israel \\
$^3$Institute for Theoretical Physics, Goethe University, 
Frankfurt, Germany 
}

\draft

\begin{document}

\maketitle

\begin{abstract}
The preliminary data on hadron multiplicities measured in central 
Pb+Pb collisions at 158$A$ GeV/$c$ are analyzed.  The ideal hadron 
gas model fails to give a reasonable explanation to the Pb+Pb data 
sets.  We study the possible effects of pion enhancement due to 
different hard-core repulsion for pions and other hadrons and 
strangeness suppression because of incomplete chemical equilibrium.  
Each of these two modifications improves the results.  The combined 
effect of these two mechanisms leads to an extremely good agreement 
with the data.  An interpretation of the obtained results in terms 
of the possible quark-gluon plasma formation at the early stage of 
the collision is also discussed.  
\end{abstract}

\pacs{PACS number(s): 25.75.Dw, 24.10.Pa}

\section{Introduction}

Different versions of thermal hadron gas (HG) models have recently 
been used to fit the data of particle number ratios in high energy 
nucleus-nucleus ($A$+$A$) collisions at BNL AGS and CERN SPS (see 
e.g.\ [1-8]).  These particle number ratios were measured in fixed 
rapidity intervals.  Particle abundances in HG models are described 
in terms of two {\it chemical} freeze-out parameters, temperature 
$T$ and baryonic chemical potential $\mu_B$.  However, the 
restriction on hadron yields in a fixed rapidity interval causes 
some problems.  The universal energy distributions of all hadrons 
in HG models are transformed into different rapidity distributions 
for different hadron species --- the width of rapidity distribution 
is decreased with increasing particle mass.  Therefore, to calculate 
these particle number ratios measured in the fixed rapidity 
interval, additional model assumptions on collective flow effects 
and specific values of {\it thermal} freeze-out parameters are 
required.  

In the present work, we analyze the data on hadron multiplicities 
in central Pb+Pb collisions at 158$A$ GeV/$c$ measured at CERN SPS.  
The aim of our thermal HG analysis of hadron multiplicities in a 
full 4$\pi$-geometry is to avoid the above-mentioned complications.  
The basic assumption here is rather strong that the values of the 
chemical freeze-out parameters $T$ and $\mu_B$ in different elements 
of the system are equal.  However, the collective flow effects are 
still permitted and the validity of a pure thermal (``fireball'') 
model is not necessarily required.  

\section{The models}

The ideal gas equation of state for a mixture of different particle 
species $i$=1,...,$h$ with chemical potentials $\mu_1,$...,$\mu_h$ 
is given by the pressure function in the grand canonical ensemble 
as 
\begin{equation}\label{pid}
p^{id}(T,\mu_1,...,\mu_h)= \sum_i^h p_i^{id}(T,\mu_i) = 
\sum_i^h \frac{d_i}{6\pi^{2}}~ \int_{0}^{\infty}dk~ 
\frac{k^4}{(k^2+m_i^2)^{1/2}}~ f_i(k)~, 
\end{equation}
where the integration variable $k$ is the momentum, and $d_i$ and 
$m_i$ are the degeneracy factor and the mass of the $i$th hadron, 
respectively.  The distribution function is written as 
\begin{equation}\label{distr}
f_i(k)=\left[\exp\left(\frac{(k^{2}+m_i^{2})^{1/2}~-~ 
\mu_i}{T}\right)~+~\eta_i ~\right]^{-1}~.  
\end{equation}
The statistic factor $\eta_i$ is $-1$ for bosons and +1 for 
fermions.  The thermal number of $i$th particle species is 
expressed as 
\begin{equation}\label{nid}
N_i^{id}(T,\mu_i) ~ \equiv ~ V ~\frac 
{\partial\,p^{id}} {\partial \mu_i} ~=~ 
\frac{d_iV}{2\pi^{2}}~\int_{0} 
^{\infty}dk~k^{2}~f_i(k)~, 
\end{equation}
in which $V$ is the total volume of the hadron system assumed to 
be equal to the sum of the proper volume elements at the chemical 
freezeout.  Hadron chemical potentials $\mu_i$ are expressed as 
\begin{equation}\label{mui}
\mu_i~=~b_i\mu_B ~+~s_i\mu_S 
\end{equation}
in terms of baryonic ($\mu_B$) and strange ($\mu_S$) chemical 
potentials with hadron charges $b_i=0,\pm 1$ and 
$s_i=0,\mp 1,\mp 2, \mp 3$.  Direct (thermal) particle yields are 
calculated according to Eq.~(\ref{nid}) for all known stable hadrons 
and resonances with mass up to 2 GeV.  The total multiplicity of 
$i$th hadron is the sum of its direct yield (\ref{nid}) and all 
possible strong decay contributions from resonances.  This is true 
for all HG models discussed in the present paper.  

There are serious theoretical doubts in the validity of the ideal 
HG model at the chemical freezeout.  Particle number densities and 
total energy density of the HG for chemical freeze-out parameters 
found from fitting experimental data look artificially large.  
This is hardly consistent with the picture of a gas of point-like 
non-interacting hadrons.  The introduction of the ``hard-core'' 
repulsion in the HG model has been widely discussed in the 
literature (e.g., [1-8]).  We follow the excluded volume procedure 
of Ref.~\cite{ris91} with the assumption of substituting the total 
volume $V$ by $V-\sum_i^hv_iN_i$.  The proper volume of each hadron 
$v_i$ is expressed in terms of its ``hard-core'' radius $r_i$ as 
$v_i=\frac{16}{3}r_i^3$ \cite{grand97}.  The HG equation of state 
with Van der Waals (VDW) repulsion can then be written as 
\begin{equation}\label{pvdw}
p^{VDW}(T,\mu_1,..., \mu_h) = 
p^{id}(T,\tilde{\mu}_1,...,\tilde{\mu}_h),~~~ 
\tilde{\mu}_i~ \equiv~ \mu_i 
~ - ~v_i~ p^{VDW}(T,\mu_1,..., \mu_h)~, 
\end{equation}
and particle multiplicities are given as 
\begin{equation}\label{nvdw}
N_i^{VDW}(T,\mu_1,...,\mu_h)~\equiv~ 
V~\frac{\partial p^{VDW}}{\partial \mu_i}~ 
=~\frac{N_i^{id}(T, \tilde{\mu}_i)} 
{1~+~\sum_{j=1}^h 
v_j~N_j^{id}(T, \tilde{\mu}_j)/V}~.  
\end{equation}
All HG multiplicities (\ref{nvdw}) are suppressed in comparison with 
the ideal HG results (\ref{nid}).  The same values of $N_i$ in the 
VDW HG as in the ideal HG correspond to a larger value of volume $V$ 
in the VDW HG than in the ideal HG.  An artificially high energy 
density in the gas of non-interacting hadrons is therefore 
removed \cite{grand97,grand98}.  

The ideal HG model can not fit existing data of hadron 
multiplicities.  Numerous attempts to fit data on particle number 
ratios in $A$+$A$ collisions at AGS and SPS energies have revealed 
two possible ways to modify the ideal HG model: pion enhancement or 
strangeness suppression.  In other words, in order to fit data, one 
usually needs either more pions or fewer strange particles in 
comparison with the ideal HG results.  The introduction of the 
phenomenological parameters $r_i$ could change particle number 
ratios in comparison with ideal HG results.  To enhance pion/hadron 
ratios we assume a smaller ``hard-core'' radius for pion than those 
for all other hadrons, $r_{\pi}<r_i=r$ \cite{rit97,grand97}.  The 
alternative dynamical explanations of the observed pion excess 
are presented in Refs.~\cite{hydr,transp} in the framework of 
hydrodynamical and transport approaches.

A chemical non-equilibrium HG with a strangeness suppression was 
suggested in Ref.~\cite{raf91}.  It is usually formulated in terms 
of a phenomenological parameter $\gamma_s <1$, which can be 
connected to additional chemical potential regulating the absolute 
value of total number of strange quarks and antiquarks in the HG 
system \cite{hein95}.  The form of Eqs.\ (\ref{pid},\ref{nid}) 
remains the same in this approach, but the distribution functions 
are modified as 
\begin{equation}\label{distrg}
f_i^s(k) = 
\left[\gamma_s^{-|S_i|}\exp\left(\frac{(k^{2}+m_i^{2})^{1/2}~ 
-~ \mu_i}{T}\right)~+~\eta_i 
~\right]^{-1}~, 
\end{equation}
where $|S_i|$ is the sum of the number of strange quarks and 
antiquarks in hadron $i$.  Hence Eq.\ (\ref{nid}) becomes 
\begin{equation}\label{nids}
N_i^{s}(T,\mu_i,|S_i|) ~ 
= ~ \frac{d_iV}{2\pi^{2}}~\int_{0} ^{\infty}dk~k^{2}~f_i^s(k)~.  
\end{equation}
Hadron multiplicity data for CERN SPS nucleus-nucleus collisions 
were recently fitted with Eq.~(\ref{nids}) in Ref.~\cite{bec97}.  

Hadron multiplicities (\ref{nid},\ref{nvdw},\ref{nids}) in HG models 
depend on 3 thermodynamical parameters, $V$, $T$, and $\mu_B$ 
(strange chemical potential $\mu_S$ is always defined by the 
requirement of zero total strangeness).  The model with strangeness 
suppression (\ref{nids}) introduces an additional phenomenological 
parameter $\gamma_s$.  Meanwhile, in the VDW HG model (\ref{nvdw}), 
one new parameter $r$ is added to the ideal HG formalism --- 
$r_{\pi}=0$ is assumed for simplicity \cite{rit97,grand97}.  Note 
that, by putting $\gamma_s=1$ and $r=0$, the ideal HG model is 
recovered: Eqs.\ (\ref{nids}) and (\ref{nvdw}) are reduced to 
Eq.\ (\ref{nid}).  

We consider also the HG model which combines both strangeness 
supperession and hard-core repulsion effects.  The pressure function 
$p$ in this model is defined by the equation 
\begin{equation}\label{pssvdw}
p~=~p^{s}(T,\tilde{\mu}_1,...,\tilde{\mu}_h),~~~ 
\tilde{\mu}_i~ \equiv~ \mu_i 
~ - ~v_i~ p~, 
\end{equation}
where $p^s$ in Eq.~(\ref{pssvdw}) is given by Eq.~(\ref{pid}), but 
with distribution functions $f_i^s$ (\ref{distrg}) instead of 
$f_i$ (\ref{distr}).  Thermal hadron multiplicities $N_i$ are then 
given by 
\begin{equation}\label{nssvdw}
N_i~ 
= ~\frac{N_i^{s}(T, \tilde{\mu}_i,|S_i|)} 
{1~+~\sum_{j=1}^h 
v_j~N_j^{s}(T, \tilde{\mu}_j, |S_i|)/V}~.  
\end{equation}

\section{Fitting procedure and results}

The data for hadron multiplicities in Pb+Pb collisions at 158$A$ 
GeV/$c$ and references to the original papers [16--21] are presented 
in the last two columns of Table I.  Because the measurement of 
hadron multiplicities in Pb+Pb collisions at SPS energies are not 
yet complete, we also include several hadron ratios in the central 
rapidity region, K$^+$/K$^-$, $\overline{\mbox{p}}/\mbox{p}$, 
$\overline{\Xi}/\Xi$ and $\overline{\Omega}/\Omega$ (the 
controversial $\overline{\Lambda}/\Lambda$ ratio is ignored) in our 
fitting.  Due to equal mass, the kinematic complications mentioned 
in previous paragraphs are not expected to be very important.  We 
assume, therefore, that these ratios in the central rapidity region 
are approximately equal to the corresponding ratios of the total 
hadron multiplicities.  

The fitting procedure is to minimize 
\begin{equation}\label{chi}
\chi^2~=~ \sum_{k=1}^N \frac{(y_k^{exper} - y_k^{theor})^2} 
{(\Delta y_k^{exper})^2}~.  
\end{equation}
To compare the fitting of different models to the data, we normalize 
$\chi^2$ by 
\begin{equation}
\frac{\chi^2}{\mbox{dof}}~\equiv~\frac{\chi^2}{N-n}~, 
\end{equation}
where $N$ is the number of experimental data points fitted and $n$ 
is the number of parameters in each model.  The best fit with ideal 
HG model is presented in the first column (ID) in Table I.  It has 
been known that the ideal HG model can not fit all hadron ratios 
simultaneously.  This problem has been discovered in the previous 
attempts to fit data on particle number ratios in collisions of S 
nuclei with different targets at CERN SPS \cite{cleym93b,cleym94}.  
From our fit of the data set in Table I for Pb+Pb it follows that 
the main problems for the ideal HG model are indeed a deficiency 
of pions, the major contribution to $h^-$, and a surplus of strange 
particles --- K$_s^0$ and $\phi$ substantially exceed their 
experimental estimates.  It is possible to improve the agreement 
with $h^-$ and $\phi$ within the ideal HG model ($h^-$=603, 
$\phi$=6.38) by choosing a smaller temperature ($T=125$ MeV) and a 
larger volume ($V=15900$ fm$^3$).  To keep $N_B$ close to data 
($N_B$=370) one needs then a larger value of the baryonic chemical 
potential ($\mu_B=300$ MeV).  These ``small'' $T$ and ``large'' 
$\mu_B$ parameters in ID HG lead, however, to completely wrong 
antibaryon to baryon ratios: $\overline{\mbox{p}}/\mbox{p}$, 
$\overline{\Xi}/\Xi$ and $\overline{\Omega}/\Omega$ all become an 
order of magnitude smaller than their measured values.  

The agreement is improved if one adopts either the strangeness 
suppression scheme (SS) or hadron ``hard-core'' repulsion scheme 
(VDW) with $r_{\pi}=0$ and $r>0$ which leads to the pion enhancement 
as mentioned before.  The VDW results are shown in the second column 
of Table I.  Note that $h^-$ becomes closer to data, but K$^0_s$ and 
$\phi$ are still too large.  The SS results are shown in the third 
column of Table I.  The quality of the fit is significantly 
improved.  However, there is still a deficiency of pions in the SS 
HG model.  Finally, the results of taking both SS and VDW together 
into consideration are given in the fourth column of Table I.  The 
$\chi^2$ of this fit is remarkably small as shown in the table.  
In addition, we present in Table II all relevant total hadron 
multiplicities calculated with parameters found from SS+VDW.  
Similar to Table I the hadron multiplicities presented in Table II 
include both direct (thermal) yield (\ref{nssvdw}) and all possible 
contributions from strong resonanse decays.  

\section{Discussion}

Let us discuss a possible physical interpretation of the obtained 
results in terms of the quark-gluon plasma (QGP) formation at the 
early stage of $A$+$A$ collisions.  The ``enhancement'' in strange 
hadron productions was suggested as a signal for the formation of 
QGP in $A$+$A$ collisions (see e.g.\ Ref.~\cite{koch86}).  One 
expects that the strangeness equilibration time in the QGP is 
much shorter than in the HG.  Therefore, rapid production and 
equilibration of strangeness is expected in the QGP initial state, 
and the strangeness suppression factor ($\gamma_s<1$) is usually 
associated with an incomplete strangeness equilibrium in the pure 
hadron matter.  

A real picture could be, however, very different.  We found that, 
for the chemical freeze-out parameters $T$ and $\mu_B$ shown 
in Table I, the strangeness to entropy ratio is larger in the 
equilibrium HG than in the equilibrium QGP.  Therefore, $\gamma_s<1$ 
strangeness {\it suppression} in the HG would become a signal for 
the formation of QGP at the early stage of $A$+$A$ collisions at 
CERN SPS energies.  The same conclusion was obtained in 
Ref.~\cite{gaz98}.  

Strange quark-antiquark pairs may be primarily produced by hard 
(with typical momenta of several GeV) (anti)quark and gluon 
collisions at the early non-equilibrium stage of $A$+$A$ at the 
SPS.  If the QGP is formed one can expect the complete strangeness 
equilibration ($\gamma_s^Q=1$ in the QGP).  To estimate the 
strangeness/entropy ratio in the QGP, let us use the ideal gas 
approximation of massless {\it u}-, {\it d}-(anti)quarks, gluons and 
strange (anti)quarks with $m_s=150$ MeV.  In the wide range of $T$ 
and $\mu_B$ ($T=160$\,--\,300 MeV, $\mu_B=0$\,--\,400 MeV), one 
finds an almost constant value of the ratio of total number of 
strange quarks and antiquarks $N_s+N_{\bar{s}}$ to the total entropy 
$S$: 
\begin{equation}\label{sentr}
R_s~\equiv~\frac{N_s+N_{\bar{s}}}{S}~=~0.022 \mbox{\,--\,} 0.025~.  
\end{equation}

There is no realistic model for the QGP hadronization.  We do not 
expect, however, essential additional contribution to the 
strange-antistrange {\it hadron} production during the QGP 
hadronization: at this stage the typical thermal (anti)quark and 
gluon momenta are only a several hundred of MeV.  Therefore, we 
assume the same number of strange quarks and antiquarks in the QGP 
before hadronization and in the HG after hadronization.  This number 
is not expected to be further changed in the HG with local thermal 
equilibrium.  The typical thermal hadron momenta are too small for 
strange hadron production.  We expect, therefore, the same number of 
$N_s+N_{\bar{s}}$ in the HG at the chemical freezeout.  The total 
entropy is also expected to be conserved approximately during the 
hadronization of QGP and HG expansion.  This suggests that the value 
of $R_s$ at the HG chemical freezeout should be close to that in the 
QGP.  The value of $R_s$ for different HG model fits are 0.045 (ID), 
0.037 (VDW),  0.029 (SS) and 0.026 (SS+VDW).  It is remarkable that 
our best HG model fit, SS+VDW, where both pion enhancement 
($r_{\pi}=0$, $r>0$) and strangeness suppression ($\gamma_s<1$) are 
included, leads to the value of $R_s$ very close, indeed, to the QGP 
estimate (\ref{sentr}).  

\section{Summary}

In summary, we analyzed new Pb+Pb data at CERN SPS on 
hadron multiplicities in a full 4$\pi$-geometry and on 
antiparticle/particle ratios in the central rapidity region within 
various HG models.  The parameters $T$ and $\mu_B$ are found from 
fitting the data and appear to be the same, $T=$165~MeV and 
$\mu_B$=235~MeV, within four different versions of the HG model.  
The ideal HG model fails to give a reasonable explanation to the 
Pb+Pb data sets.  Meanwhile, the obtained particle number density 
and total energy density of the ideal HG at the chemical freezeout 
is extremely large, which is inconsistent with the picture of a gas 
of point-like non-interacting hadrons.  To improve this situation, 
the hadron ``hard-core'' repulsion had been proposed.  It removes 
the artificially high energy and particle number densities in the 
gas of non-interacting hadrons.  The VDW HG model with $r_{\pi}=0$ 
and $r_i=r>0$ leads to the pion enhancement and better fit of hadron 
multiplicities is obtained.  Chemical non-equilibrium effects with 
strangeness suppression were also studied.  We have found that this 
modification significantly improves the agreement with data.  
Furthermore, the combined effect of these two mechanisms results in 
an extremely good agreement with the data analyzed.  We present the 
SS+VDW HG model predictions for relevant hadron multiplicities in 
central Pb+Pb at the SPS.  Finally, an interpretation of the 
obtained results in terms of the possible QGP formation at the early 
stage of $A$+$A$ collisions at the SPS is also discussed.  

\acknowledgments
The authors are thankful to Marek Ga\'zdzicki for useful comments 
and to Horst St\"ocker for critical remarks.  This work is supported 
in part by National Science Council of Taiwan under grand 
nos.\ NSC 87-2112-M-001-005 and 88-2112-M-001-016.  MIG gratefully 
acknowledges the support and hospitality of the School of Physics 
and Astronomy, Tel Aviv University, Israel and the Institute for 
Theoretical Physics, Goethe University, Frankfurt, Germany.  GDY 
would also like to acknowledge the use of computing facility in 
National Center for High-performance Computing of Taiwan.  

\begin{table}
\caption{ Particle multiplicities and ratios for central Pb+Pb SPS 
collisions in different HG models. }

\begin{tabular}{lrrrrrc}
                 & ID    & VDW    & SS    & SS+VDW  & Data
                                                    & Reference   \\
                                                    \tableline
$T$ [MeV]        & 165   & 165    & 165   & 165     &
                                                    &             \\
$\mu_B$ [MeV]    & 235   & 235    & 235   & 235     &
                                                    &             \\
$V$ [fm$^3]$     & 2490  & 6060   & 3150  & 6640    &
                                                    &             \\
$r$ [fm]         & 0.00  & 0.46   & 0.00  & 0.45    &
                                                    &             \\

$\gamma_s$       & 1.00  & 1.00   & 0.55  & 0.62    &
                                                    &             \\
                                                    \tableline
$N_B$            & 347   & 337    & 382   & 374     & 372$\pm$10
                                                    & \cite{dat1} \\
$h^-$            & 514   & 615    & 566   & 674     & 680$\pm$50
                                                    & \cite{dat2} \\
K$_s^0$          & 89.3  & 86.4   & 60.5  & 65.3    & 68$\pm$10
                                                    & \cite{dat2} \\
$\phi$           & 11.7  & 11.3   & 4.46  & 5.42    & 5.4$\pm$0.7
                                                    & \cite{dat3} \\
$\mbox{p}-\overline{\mbox{p}}$ 
                 & 124   & 120    & 155   & 149     & 155$\pm$20
                                                    & \cite{dat2} \\
K$^+$/K$^-$      & 1.78  & 1.78   & 1.74  & 1.75    & 1.8$\pm$0.1
                                                    & \cite{dat4} \\
$\overline{\mbox{p}}/\mbox{p}$ 
                 & 0.060 & 0.060  & 0.059 & 0.060   & 0.07$\pm$0.01
                                                    & \cite{dat5} \\
$\overline{\Xi}/\Xi$
                 & 0.242 & 0.244  & 0.223 & 0.227   &0.249$\pm$0.019
                                                    & \cite{dat6} \\
$\overline{\Omega}/\Omega$
                 & 0.496 & 0.502  & 0.441 & 0.452   &0.383$\pm$0.081
                                                    & \cite{dat6} \\
                                                    \tableline
$\chi^2$/dof     & 108/6 & 95.4/5 & 12.5/5 & 3.58/4 &   &
\end{tabular}

\vspace{2cm}

\caption{ Particle multiplicities for central Pb+Pb SPS collisions 
predicted in SS+VDW HG model.  The values of model parameters are 
taken from Table I. }

\begin{tabular}{rrrrrrrrr}
$\pi$ & K$^+$ & K$^-$ & $\Lambda$ & $\overline{\Lambda}$ &
$\Xi$ & $\overline{\Xi}$ & $\Omega$ & $\overline{\Omega}$ \\
1850 & 83.1 & 47.6 & 38.9  & 4.43 & 5.86 & 1.33 & 0.242 & 0.110
\end{tabular}

\end{table}

\end{document}